\documentclass[epj]{svjour}
\usepackage{graphics}
\usepackage{amsmath}
\usepackage{amssymb}

\usepackage{xcolor}
\begin{document}
\title{Updated Bounds on the Minimal Left-Right Symmetric Model from LHC Dilepton Resonance Searches }
\author{Gabriela Lichtenstein\inst{1} 
\and M. J. Neves\inst{2} \and Farinaldo Queiroz\inst{1,3,4,5} \and Ricardo C. Silva\inst{1,3,6}
}                     
%
%
\institute{International Institute of Physics, Universidade Federal do Rio Grande do Norte,  59078-970, Natal, RN, Brasil \and Departamento de F\'isica, Universidade Federal Rural do Rio de Janeiro,
BR 465-07, 23890-971, Serop\'edica, RJ, Brasil \and Departamento de F\'isica, Universidade Federal do Rio Grande do Norte, 59078-970, Natal, RN, Brasil
\and Departamento de F\'isica, Facultad de Ciencias, Universidad de La Serena,
Avenida Cisternas 1200, La Serena, Chile \and Millennium Institute for Subatomic Physics at High-Energy Frontier (SAPHIR), Fernandez Concha 700, Santiago, Chile \and Laboratoire de physique nucléaire et des hautes énergies (LPNHE), Sorbonne Université, Paris, France.}
%
\date{Received: date / Revised version: date}
%
\abstract{
The Left-Right model is a popular extension of the Standard Model that features the new gauge bosons, $W^{\pm}_R$ and  $Z_R$. Collider searches for a Left-Right symmetry are often concentrated on the charged right-handed current. However, in this work, we take advantage of the dilepton data at the LHC with center-of-mass energy of 13 TeV and 139 fb$^{-1}$ of integrated luminosity to place lower mass bounds in the $Z_R$ mass based on the $p\,p \rightarrow Z_{R} \rightarrow \ell^{+} \, \ell^{-}+X$ process. We vary the $SU(2)_R$ gauge coupling from $g_{R}=0.4$ to $g_{R}=1.0$, and find that the LHC data impose $M_{Z_R}>5.4$ TeV and $M_{Z_{R}}>6.1$ TeV, respectively. Lastly, we put our findings into perspective with $W_R$ searches at the LHC and show that our limits cover an unexplored region of parameter space, where the right-handed neutrino is heavier than the $W_R$ boson. 
\PACS{
      {PACS-key}{discribing text of that key}   \and
      {PACS-key}{discribing text of that key}
     } 
} 
\titlerunning{Updated Bounds on the Minimal Left-Right Symmetric Model}
\authorrunning{Gabriela Lichtenstein et al.}
\maketitle
\section{Introduction}
\label{intro}
\label{sec:intro}
The left-right symmetric model (LRSM) is a well-known extension of the Standard Model (SM). It was initially proposed to explain the parity violation in the SM \cite{PatiPRD1974,MohapatraPRD1975,PhysRevD.11.566,SenjanoviPRD1975}.  Still, it also naturally incorporates seesaw mechanisms \cite{seesaw1,seesaw2,seesaw3,seesaw4,seesaw5}, and dark matter via an extra gauge symmetry $U(1)_{X}$ portal
\cite{MJNevesHelayelMohapatraOkada2018,MJNevesOkada2021}. In the minimal version, the LRSM contains a rich sector of scalar fields that introduces three vacuum expectation values (VEVs) needed to yield masses to all matter content, including fermions, gauge bosons, and scalars. 

The LRSM has a rich phenomenology, including studies at colliders \cite{Fuks:2025jrn,ThomasArun:2021rwf,Hong:2023mwr,Liu:2025ldf,Lindner:2016lpp,De:2024puh,Nemevsek:2011hz,Nemevsek:2023hwx}, as well as tests of CP violation \cite{Zhang:2007da}, searches for lepton flavor violation \cite{Bajc:2009ft}, and neutrino physics \cite{Alves:2022yav,Maiezza:2010ic} .
One of the main signatures of the left-right symmetry is the $W_R$ boson. It has a right-handed current with the charged leptons. The production of $W_{R}$ is typically explored through the process $pp \to W_R \to N_{R} \, \ell \to lljj$, with the $W_R$ decaying into $N_R,\ell$ assuming $M_{W_R} > M_{N_R}$, where $N_R$ is the right-handed neutrino \cite{Frank:2023epx}.  Having in mind that $g_L$ and $g_R$ are the gauge couplings of the $SU(2)_L$ and $SU(2)_R$ groups, it is also worth investigating the setups where $g_R \neq g_L$ \cite{Patra:2015bga,Osland:2020onj,Solera:2023kwt,Kriewald:2024cgr}. As the right-handed neutrino can be heavier than the $W_R$ boson with no prejudice, it is worth exploring this parameter space. 
 
That said, in this work, we focus on the $Z_R$ boson instead of the $W_R$ boson. As the masses of the $Z_R$ and $W_R$ bosons are connected via $g_R$ and the scale of symmetry breaking of the left-right symmetry, any bound found on $Z_R$ can be directly translated to a lower mass limit on $W_R$ even if $M_{N_R}$ is heavier than the $W_R$ boson. Moreover, we depart from the usual $g_R=g_L$ assumption, having in mind that different realizations of the left-right symmetry with $g_R \neq g_L$ have been put forth \cite{Chakrabortty:2016wkl,Deppisch:2015cua}. This work is an update to the results in \cite{Patra:2015bga}, which analyses the LHC data at 8 TeV and ${\cal L}= 20 $ fb $^{-1}$, and the study at reference \cite{Lindner:2016lpp} at 13 TeV and ${\cal L }=  3.2$ fb $^{-1}$. We find that the updated data with 13 TeV and integrated luminosity ${\cal L }=  139$ fb $^{-1}$ has pushed the limits on the $Z_R$ mass about 2 TeV higher than the 3.2 fb $^{-1}$ data, as shown in table \ref{tab:bounds}.

We place lower mass limits on the $Z_R$ gauge boson using dilepton data from LHC Run II, with 139 fb$^{-1}$ integrated luminosity. We simulate the $p\,p \rightarrow Z_{R} \rightarrow \ell^{+} \, \ell^{-} + X$ process with center-of-mass energy of $\sqrt{s} = 13$ TeV and determine which $Z_R$ mass is consistent with the data for a given $g_R$ coupling. Our paper complements the existing literature for the following reasons: {\bf (i)} We use updated dilepton limits with $\mathcal{L}=139fb^{-1}$;  {\bf (ii)} We perform our analysis for $g_R \neq g_L$ taking into account the impact of an existing pole in the cross-section;  {\bf (iii)} We cover a different parameter space of the usual $W_R \times N_R$ parameter space.
 Moreover  dijet constraints from $pp \rightarrow Z_R \rightarrow jj$ are  disconsidered in this work. Although the production rate is higher than di-leptons, the background is too high, and di-leptons produce a clearer signal. For instance, reference \cite{ATLAS:2020zzb} produce lower bounds on $Z'$ at $\approx 2$ TeV, while di-leptons usually produce bounds at $\approx 4$ TeV \cite{ATLAS:2019erb}.

 The paper is organized as follows: Section (\ref{sec:model}) provides a brief review of the LRSM. In the section (\ref{sec:3}), we show the physical eigenstates of the new gauge bosons, couplings with fermions, and the discussion of $ Z_R$ decay width into fermion-anti-fermion pair in terms of the new gauge coupling $g_R$. In the section (\ref{sec:pheno}), we discuss the phenomenology of the process $p\,p \rightarrow Z_{R} \rightarrow \ell^{+} \, \ell^{-}$ at the LHC, with $\sqrt{s}=13$ TeV and luminosity $139\, \mbox{fb}^{-1}$. We present the final comments in the section (\ref{conc}). In the appendix (\ref{secAppendix}), we provide a detailed description of the gauge sectors of the LRSM, including diagonalization and physical eigenstates.
 
\section{Review on the minimal left-right symmetric model}
\label{sec:model}
The LRSM is based on the gauge symmetry $SU(2)_L \times SU(2)_R \times U(1)_{B-L}$, where $B-L$ is the preserved charge from the barionic (B) and the leptonic (L) numbers. For a review of the model, see the reference \cite{Lee:2017mfg}. The covariant derivative operator of the model is
\begin{eqnarray}\label{Dmu}
D_{\mu} = \partial_{\mu}-i g_{L} \, A_{L\mu}^{a} \frac{\sigma_{L}^{a}}{2}
-i g_{R} \, A_{R\mu}^{a} \frac{\sigma_{R}^{a}}{2} - i g_{BL} \, \frac{Q_{BL}}{2} B_{\mu} \; , \;\;\;
\end{eqnarray}
where $g_L$, $g_R$, and $g_{BL}$ are the coupling constants, $A_{L\mu}^{a} \,(a=1,2,3)$, $A_{R\mu}^{a}$ and $B_{\mu}$ are the gauge fields of $SU(2)_L$, $SU(2)_R$ and $U(1)_{B-L}$, respectively. The fermion sector is set by
\begin{eqnarray}\label{Lfermions}
{\cal L}_{f}=
 \overline{\psi_L^i} \, i \, \gamma^\mu D_\mu \psi_L^i
+\overline{\psi_R^i} \, i \, \gamma^\mu D_\mu \psi_R^i 
\nonumber \\
+\overline{Q_L^i} \, i \, \gamma^\mu D_\mu Q_L^i
+\overline{Q_R^i} \, i \, \gamma^\mu D_\mu Q_R^i \; ,
\end{eqnarray}
in which the content is assigned by
\begin{eqnarray}
Q_{L}^{i} &=  \left(\begin{array}{c}u_L\\d_L \end{array}\right)_i ,  \left({ \bf 3}, {\bf 2}, {\bf 1}, +\frac{1}{3}\right) \; , \;
\nonumber \\
Q_{R}^{i} &=  \left(\begin{array}{c}u_R\\d_R \end{array}\right)_i ,  \left({ \bf 3}, {\bf 1}, {\bf 2}, +\frac{1}{3}\right) \; ,
\nonumber \\
\psi_{L}^{i}  &=   \left(\begin{array}{c}\nu_L \\ \ell_L \end{array}\right)_i , \left({ \bf 1}, {\bf 2}, {\bf 1}, -1 \right) \; , \;
\nonumber \\
\psi_{R}^{i} &=  \left(\begin{array}{c} N_R \\ \ell_R \end{array}\right)_i , \left({ \bf 1}, {\bf 1}, {\bf 2}, -1 \right) \; ,
\label{lrSM}
\end{eqnarray}
where $i=1, 2, 3$ is the generation index. 
The gauge bosons' kinetic terms are given by
\begin{equation}\label{Lgauge}
{\cal L}_{gauge}=-\frac{1}{2} \,\mbox{tr}\left(F_{L\mu\nu}^{\; 2}\right)
-\frac{1}{2} \,\mbox{tr}\left(F_{R\mu\nu}^{\; 2}\right)
-\frac{1}{4} \, B_{\mu\nu}^{\; 2} \; ,
\end{equation}
where $F_{L\mu\nu}=\partial_{\mu}A_{L\nu}-\partial_{\nu}A_{L\mu}+i\,g_{L}\left[ \, A_{L\mu} \, , \, A_{L\nu}\right]$, $F_{R\mu\nu}=\partial_{\mu}A_{R\nu}-\partial_{\nu}A_{R\mu}+i\,g_{R}\left[\,A_{R\mu} \, , \, A_{R\nu} \, \right]$ and $B_{\mu\nu}=\partial_{\mu}B_{\nu}-\partial_{\nu}B_{\mu}$ 
are the field-strength tensors of $A_{L\mu}^{\, a}$, $A_{R\mu}^{\, a}$ and $B_{\mu}$, respectively. 

In the minimal version of LRSM, the scalar sector is governed by the kinetic term
\begin{eqnarray}\label{LScKin}
{\cal L}_{Sc}^{Kin}= \mbox{Tr}(|D_{\mu}\Phi|^2)
+\mbox{Tr}(|D_{\mu}\Delta_{L}|^2)
+\mbox{Tr}(|D_{\mu}\Delta_{R}|^2) \; ,
\end{eqnarray}
where multiplets set the scalar fields:
\begin{align}
\Phi &= \left(\begin{array}{cc}\phi^0_1 & \phi^+_2\\\phi^-_1 & \phi^0_2\end{array}\right) : ({\bf 1}, {\bf 2}, {\bf 2}, 0)
\; , \; \nonumber\\
\Delta_L &= \left(\begin{array}{cc}\Delta^+_L/\sqrt{2} & \Delta^{++}_L\\\Delta^0_L & -\Delta^+_L/\sqrt{2}\end{array}\right) : ({\bf 1}, {\bf 3}, {\bf 1}, 2)
\; , \; \nonumber \\
\Delta_R &= \left(\begin{array}{cc}\Delta^+_R/\sqrt{2} & \Delta^{++}_R\\\Delta^0_R & -\Delta^+_R/\sqrt{2}\end{array}\right) : ({\bf 1}, {\bf 1}, {\bf 3}, 2) \; .
\label{eq:scalar}
\end{align}

The electric charge of the model is defined by
\begin{eqnarray}
\label{Qem}
 Q_{em}=I_{3L}+I_{3R}+\frac{Q_{BL}}{2} \; ,
\end{eqnarray}
where $I_{3L(R)}=\sigma_{L(R)}^{3}/2=\pm 1/2$ are the isospins of $SU(2)_{L}$ and  $SU(2)_{R}$, respectively, and $Q_{BL}$ is the generator of $U(1)_{B-L}$. The symmetry is broken by the vacuum expectation value (VEV) scales
\begin{align}\label{VEVs}
   \langle \Phi \rangle &= \begin{pmatrix}
\kappa_1/\sqrt{2} & 0 \\
0 & \kappa_2 \, e^{i\alpha}/\sqrt{2}
\end{pmatrix} 
\; , \;   \nonumber \\
\langle \Delta_L \rangle &= \begin{pmatrix}
0 & 0 \\
v_L \, e^{i\theta_L}/\sqrt{2} & 0
\end{pmatrix} 
\; , \;   \nonumber \\
\langle \Delta_R \rangle &= \begin{pmatrix}
0 & 0 \\
v_R/\sqrt{2} & 0
\end{pmatrix} \; , 
\end{align}
for two real phases $\alpha$ and $\theta_{L}$, that obeys the hierarchy condition $v_{R} \gg v=\sqrt{ \kappa_1^2+\kappa_2^2 } \gg v_{L}$, with $\kappa_1=v \sin \beta$ and $\kappa_2=v \cos \beta$, in which $v=246 \, \mbox{GeV}$ is the VEV scale of the SM. After the SSB, the neutrinos acquire masses via the type II seesaw mechanism.
%
%

%
\section{Physical gauge bosons and couplings in LRSM}
\label{sec:3}
%
%

The charged gauge bosons (CGB) sector can be written in matrix form:
\begin{align}
{\cal L}_{mass}^{CGB}&=\frac{1}{2}
\begin{pmatrix}
A_L^- & A_R^-
\end{pmatrix}
M_{C}
\begin{pmatrix}
A_L^+ \\ A_R^+
\end{pmatrix} \; , 
\nonumber \\
M_{C} &= \begin{pmatrix}
( \, g_L^2 \, v^2 + 2 g_L^2 \, v_L^2 \, )/2 & -\,e^{i \alpha} \, g_L \, g_R \, v^2 \, \sin(2\beta)/2 \\
- e^{-i \alpha} \, g_L \, g_R \, v^2 \, \sin(2\beta)/2 
& \frac{1}{2} \, g_R^2 \, v^2 + g_R^2 \, v_R^2
\end{pmatrix}
\end{align}
where $A_{L,R}^\pm = (A^1_{L,R} \mp i \, A^2_{L,R})/\sqrt{2}$. A unitary transformation can diagonalize the mass matrix (see the appendix \ref{secAppendix}), and using the hierarchy condition of the VEVs, the masses of the physical bosons $W_{L}^{\pm}$ and $W_{R}^{\pm}$ are given by  
\begin{eqnarray}
\label{massesWW_{R}}
M_{W_{L}} \simeq \frac{ g_{L} \, v }{2}
\hspace{0.5cm} \mbox{and} \hspace{0.5cm}
M_{W_{R}} \simeq \frac{ g_{R} \, v_R}{\sqrt{2}} \; ,
\end{eqnarray}
respectively. In the gauge neutral bosons (NGB) sector, the matrix form for the massive lagrangian is
\begin{align}\label{LmassZ}
{\cal L}_{mass}^{NGB}&=
\frac{1}{2} \, 
\left(
\begin{array}{ccc}
A_{L\,\mu}^{\,\,3} & A_{R\,\mu}^{\,\,3} & B_{\mu} \\
\end{array}
\right)
M_N
 \left(
\begin{array}{c}
A_{L}^{\,\,  \mu \, 3} \\
A_{R}^{\,\,  \mu \, 3} \\
B^{\mu} \\
\end{array}
\right) \; , \nonumber \\
M_N &=  \left(
\begin{array}{ccc}
\frac{1}{4}\,g_{L}^2\,(v^2+4v_L^2) & -\frac{v^2}{4}\,g_{L}\,g_{R} & -g_L\,g_{BL}\,v_{L}^2
\\
-\frac{v^2}{4}\,g_{L}\,g_{R} & v_{R}^{2}\,g_{R}^{2}+\frac{v^2g_{R}^{2}}{4} & -g_{L}\,g_{BL}v_{R}^2
\\
-g_{L}\,g_{BL}\,v_{L}^2 & -g_{R}\,g_{BL}\,v_{R}^2 & g_{BL}^{2}\,(v_L^2  + v_{R}^2)
\end{array}
\right) \; ,
\end{align}
that can be diagonalized by a $SO(3)$ transformation that introduces three mixing angles $(\theta_{1},\theta_{2},\theta_{3})$ (see the appendix \ref{secAppendix}). The diagonalized matrix from (\ref{LmassZ}) is $M_{D}^2=\mbox{diag}(M_{Z}^2,M_{Z_{R}}^2,0)$ whose eigenvalues are read as
\begin{align}
M_{Z}&\simeq\frac{v}{2}\,\sqrt{ \, \frac{g_{L}^2 \, g_{R}^2+g_{L}^2\,g_{BL}^2+g_{R}^2\,g_{BL}^2 }{g_{R}^2+g_{BL}^2} \, }  
\hspace{0.5cm} \mbox{and} \hspace{0.5cm}  \nonumber \\
M_{Z_{R}}&\simeq v_{R} \, \sqrt{g_{R}^2+g_{BL}^2} \; ,
\label{eq:MZR_vR}
\end{align}
for the condition of $v_{R} \gg v \gg v_{L}$. The null eigenvalue corresponds to the photon mass, and the physical eigenstate associated with it is identified as the electromagnetic field. The mixing angles satisfy the parameterization for the fundamental charge $(e)$
\begin{eqnarray}\label{parametrization}
e=g_{L}\,\sin\theta_2=g_{R}\,\sin\theta_{1}\,\cos\theta_2=g_{BL}\, \cos\theta_{1}\,\cos\theta_{2} 
\; ,  \; \; \;
 \label{eq:angles}
\end{eqnarray}
where $\cos\theta_{2}=M_{W}/M_{Z}\equiv \cos\theta_{W}$ is identified as the Weinberg angle $\sin^2\theta_{W}=0.23$. From (\ref{parametrization}), the mixing angles $\theta_{1}$ and $\theta_{3}$ are given by
\begin{align}\label{theta1theta3}
\theta_{1}&=\sin^{-1}\left( \frac{g_{L}}{g_{R}} \, \tan\theta_{W} \right)  
\; \; \; \mbox{and} \; \; \; \nonumber \\
\theta_{3}&\simeq -\cos\theta_{W}\,\sqrt{ \, \frac{g_{R}^2}{g_{L}^2}-\tan^2\theta_{W} \, } \; \frac{M_{Z}^2}{M_{Z_{R}}^2} \; ,
\end{align}
that are written in terms of the new parameters $g_{R}$ and $M_{Z_{R}}$. 
Using this parametrization, the masses of the neutral gauge bosons are written as
\begin{align}
M_{Z}&=\frac{g_{L}\,v}{2\cos\theta_{W}} 
\hspace{0.5cm} \mbox{and} \hspace{0.5cm} \nonumber \\
M_{Z_{R}}&=\frac{g_{R} \, v_{R}}{ \sqrt{1-(g_{L}/g_{R})^2\,\tan^2\theta_{W}} } \; 
\label{eq:BosonsMasses}.
\end{align}
The masses of the left-handed $W$ and $Z$ correspond to the SM gauge boson masses $M_{W}=80$ GeV and $M_{Z}=91$ GeV by an appropriate choice of the parameters. The $M_{W_{R}}$ and $M_{Z_{R}}$ masses both depend on the $v_{R}$-VEV and should lie at the TeV scale. 

In terms of the physical fields, the couplings from the SM are reobtained for $Z^{\mu}$ and the photon $A^{\mu}$. The couplings of fermions with the charged gauge bosons $W_L^{\pm}$ and $W_{R}^{\pm}$ are given by
\begin{align}
{\cal L}^{int}_{CGB}=\frac{g_L}{\sqrt2} \, W_{L\mu}^{+} \, \left[ \, \bar{u}_L \, \gamma^\mu \, d_L + \bar{\nu}_L \, \gamma^\mu \, \ell_L \, \right] + 
\nonumber \\
+\frac{g_R}{\sqrt2} \, W_{R\mu}^{+} \, \left[ \, \bar{u}_R \, \gamma^\mu \, d_R + \bar{N}_R \, \gamma^\mu \, \ell_R \, \right] + {\rm h. \, c.} \; ,
\end{align}
and the couplings with the neutral gauge bosons are read as
\begin{align}
{\cal L}_{NGB}^{int} = e \, J_{\mu}^{em} A^{\mu}+\frac{g_{L}}{\cos\theta_{W}} J_{Z\,\mu}^{0} \, Z^{\mu} + 
\nonumber \\
 + g_{Z_{R}} \, J_{Z_{R} \, \mu}^{0} Z_{R}^{\, \mu} \; ,
\end{align}
where $J_{\mu}^{em}=Q_{em}^{f}\,\overline{f}_{L,R} \, \gamma_{\mu}\, f_{L,R}$ is the EM current, $J_{Z\,\mu}^{0}=J_{L \, \mu}^{3}-\sin^2\theta_{W} \, J_{\mu}^{em}$ 
is the neutral current of $Z^{\mu}$, $J_{Z_{R} \, \mu}^{0} = J_{R \, \mu}^{3}+ \left(J_{L \, \mu}^{3}  - J_{\mu}^{em} \right)(g_{L}/g_{R})^2\tan^2\theta_{W}$ 
is the neutral current of $Z_{R}^{\mu}$ with coupling constant 
\begin{equation}
    g_{Z_{R}}=\frac{g_R}{\sqrt{1 - (g_L / g_R)^2 \, \tan^2 \theta_W }} \; ,
    \label{eq:gzr}
\end{equation}
in which the isospin left- and right-currents are $J_{L(R) \, \mu}^{3}=I_{3L(R)}\,\overline{f}_{L,R}\,\gamma_{\mu}\,f_{L,R}$. 
The notation shows $f_{L, R}$ as any chiral fermion with
left- or right-handed components from the LRSM content.
Explicitly, the neutral current of $Z_{R}$ is given by 
\begin{eqnarray}
J_{Z_{R} \, \mu}^{0} = g_{L}^{\, f} \, \overline{f} \, \gamma_{\mu} \, P_{L} \, f + g_{R}^{\, f} \, \overline{f} \, \gamma_{\mu} \, P_{R} \, f \; ,
\label{JZ_{R}}
\end{eqnarray}
in which we write $f_{L(R)}=P_{L(R)}\,f$ in terms of the projectors $P_{L}=(1-\gamma_{5})/2$ and $P_{R}=(1+\gamma_{5})/2$,
and the left- and right-coefficients are

\begin{eqnarray}
    g_{L(R)}^{\, f} = I_{3R}+ \left( I_{3L}- \, Q_{em}^{\, f} \right)\tan^2\theta_{W}\,\left(\frac{g_{L}}{g_{R}}\right)^2 \ ,
    \label{eq:gf}
\end{eqnarray}
 where the charges $I_{3R}$, $I_{3L}$and $Q_{em}^f$ vary according to the left (right) representation of the f-fermion.

Using the couplings of $Z_{R}$ with fermions, the interaction Lagrangian is as follows:
\begin{align}
{\cal L}^{int}_{Z_{R}-f} = 
g_{Z_{R}}
\, Z_{R\mu} \, 
\overline{f} \, \gamma^{\mu} 
\left[ \, I_{3R} + \frac{g_L^2}{g_R^2} \, \tan^2 \theta_W \left(I_{3L} - Q_{em}^{f}\right) \, \right] f \; ,
\label{eqzprimeint}
\end{align} 
and the decay width $Z_{R} \rightarrow \bar{f} \, f$ is read as 
\begin{eqnarray}
\Gamma(Z_{R} \rightarrow \bar{f} \, f ) = N_{c}^{f} \, \frac{M_{Z_{R}}}{48\pi} \, g_{Z_{R}}^2
\left( |g^{f}_{V}|^2+|g^{f}_{A}|^2 \right) 
\nonumber \\
\times \sqrt{1-4\, \frac{m_{f}^{\, 2}}{M_{Z_{R}}^{\, 2}}} \left( \, 1
+\, \frac{2m_{f}^{\, 2}}{M_{Z_{R}}^{\, 2}} \, \right) \; ,
\label{eq:decaywidth}
\end{eqnarray}
where $N_{c}^{f}$ is the color factor for a $f$-fermion, and the mass $m_{f}$ of a $f$-fermion of the model that satisfies condition $M_{Z_{R}}>2m_f$, if the process is allowed kinetically. For leptons and neutrinos, $N_{c}=1$, and for quarks, $N_{c}=3$. The vector and axial coefficients for a $f$-fermion are defined by $g_{V/A}^{\, f}=2\left( \, g_{L}^{\, f} \pm g_{R}^{\, f} \, \right)$.
%

%
\begin{figure}
\resizebox{0.5\textwidth}{!}{%
  \includegraphics{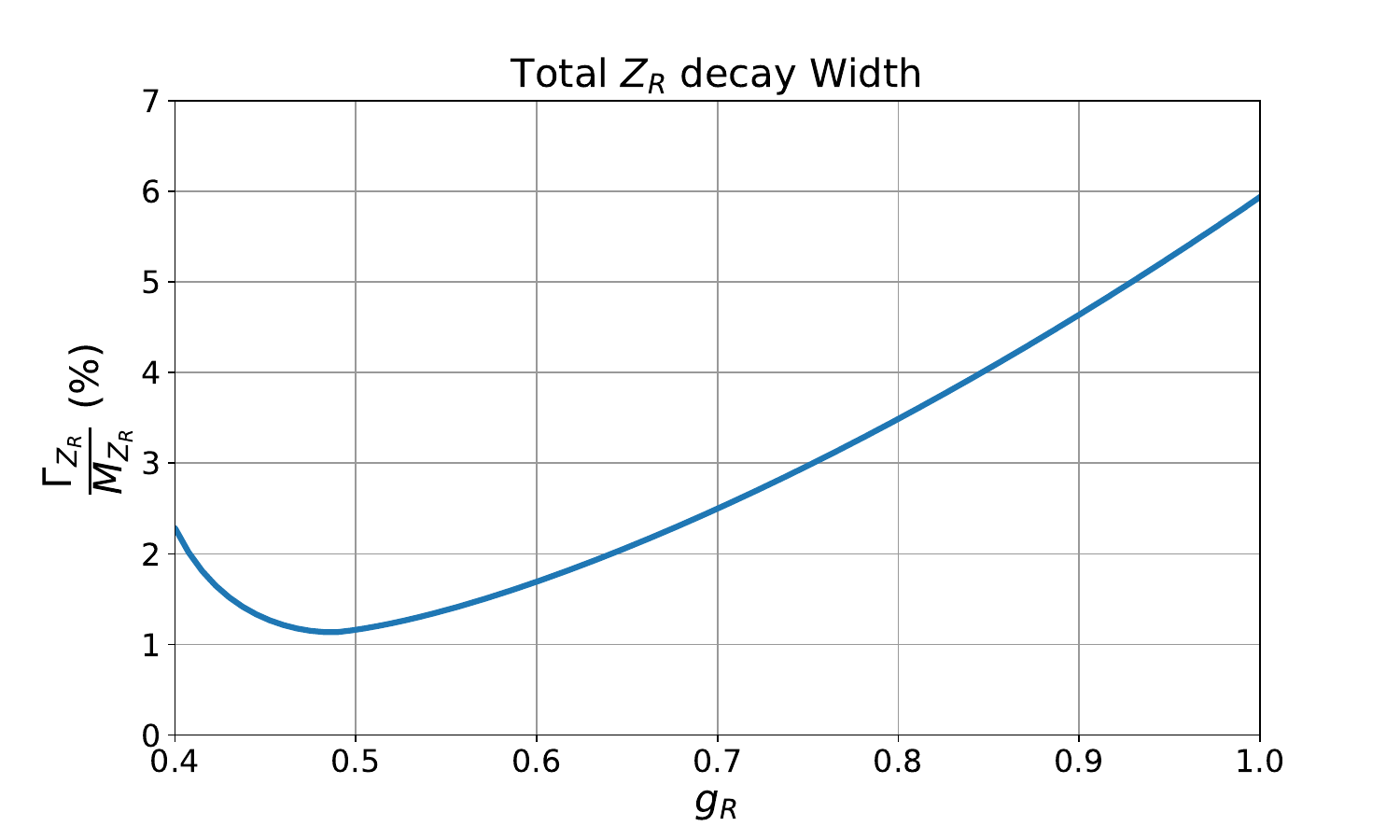}
}
    \caption{Total decay Width of  $Z_R$ divided by its mass $M_{Z_R}$ in percentage in terms of the coupling $g_R$.}
    \label{fig:decay}    
\end{figure}

Figure \ref{fig:decay} shows the total decay width of $Z_R$ divided by its mass in terms of the gauge coupling $g_R$. The decay width into fermions is described in equation \ref{eq:decaywidth}, which makes up most of the total width. The dependence of $\Gamma/M_{Z_R}$ can be considered constant with the mass $M_{Z_R}$ when neglecting terms proportional to $\frac{m_f^2}{M_{Z_R^2}} \ll 1$. The channel $Z_R \to W_R \, W_R$ is also considered when $M_{Z_R} > 2 M_{W_R}$. The full list of all $Z_R$ decay modes can be found in Table 1 of reference \cite{Solera:2023kwt}. However, we have not considered the decay channels that include the $W_L$ bosons, since they are suppressed by the charged boson mixing angle $\sin{\chi} \propto \frac{v}{v_R} \ll 1$. We are also assuming that all the physical scalars are too heavy to be produced, except $h_0$, which is identified as the Higgs with $m_{h_0} = 125$ GeV. According to Table 1 of \cite{Zhang:2007da}, the heavy scalar modes are proportional to $v_R^2$.

\section{The phenomenology at the LHC}
\label{sec:pheno}

%
\begin{figure}
\resizebox{0.47\textwidth}{!}{%
  \includegraphics{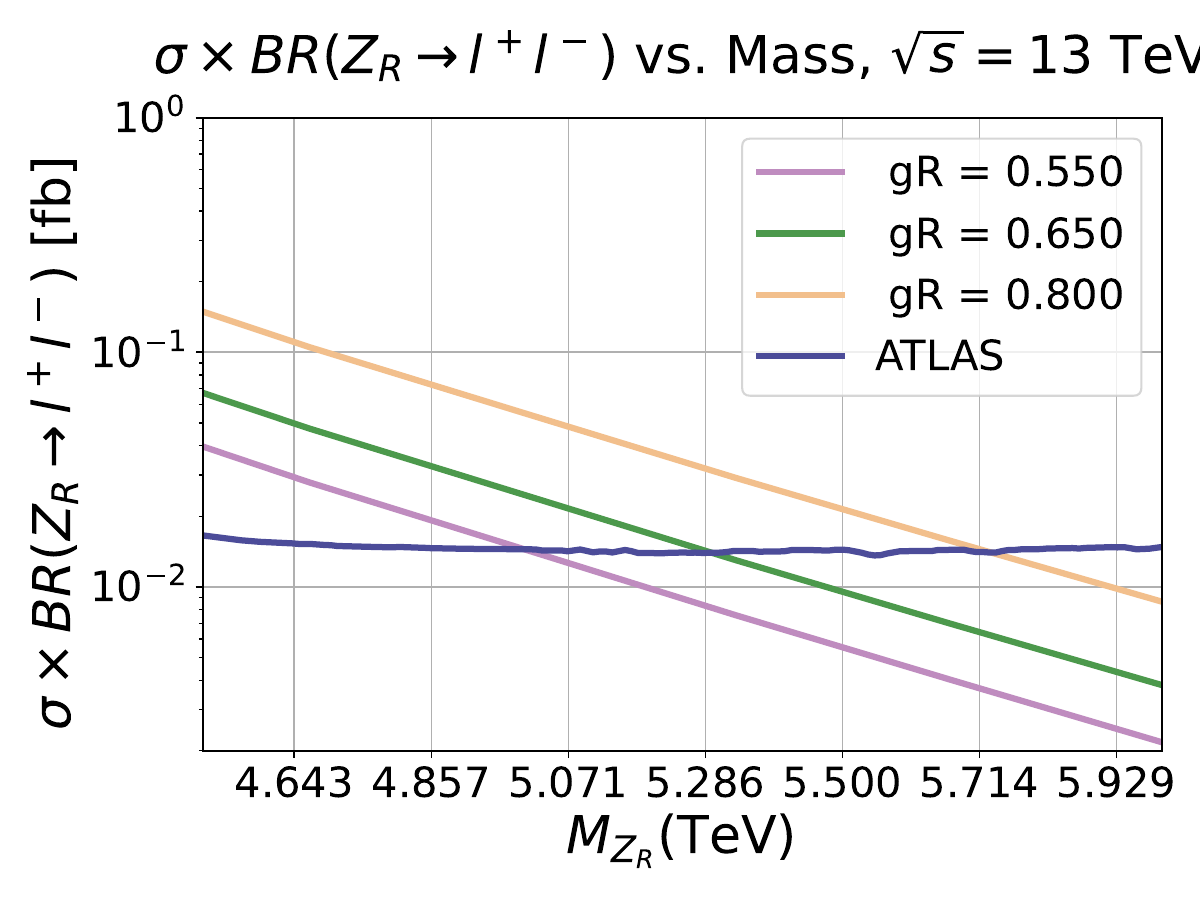}
}

\resizebox{0.47\textwidth}{!}{%
  \includegraphics{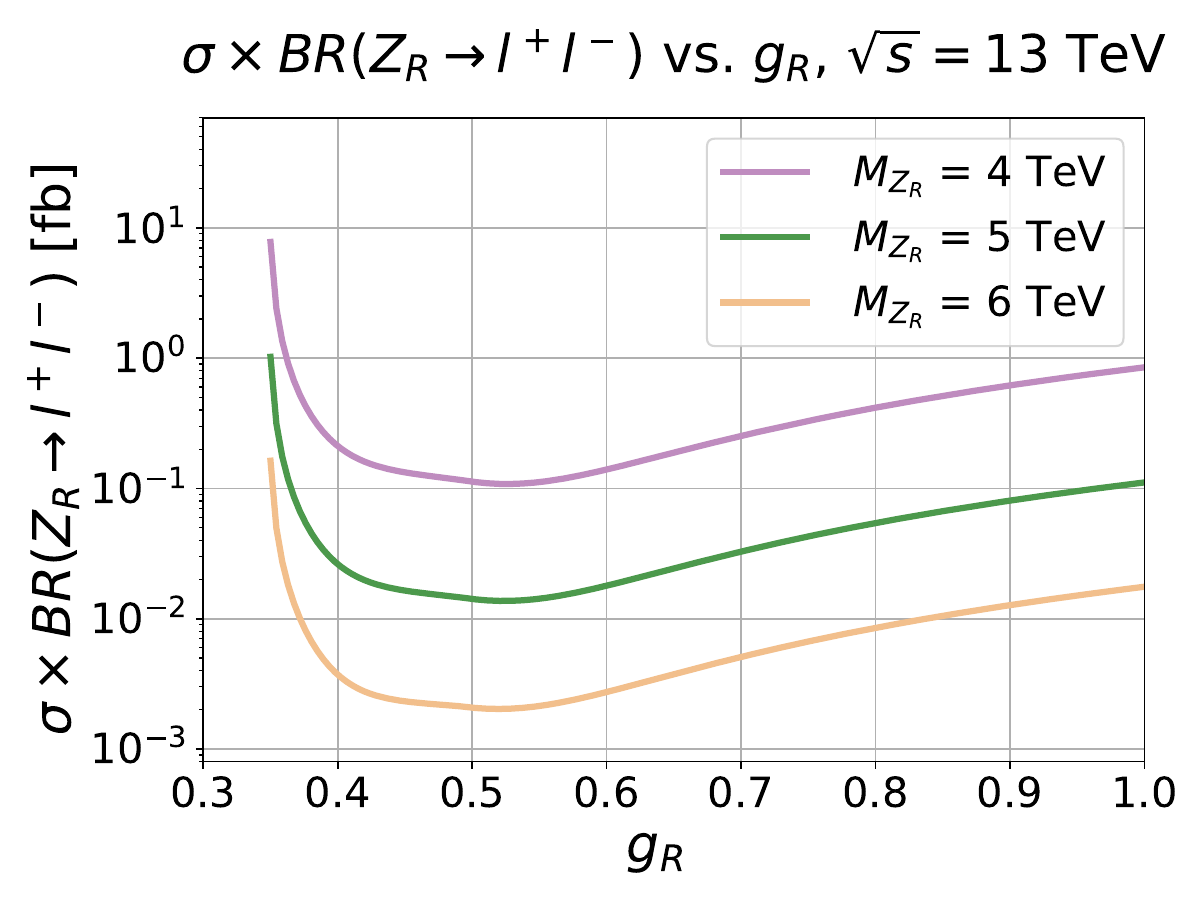}
}
    
    \caption{Cross section p p $\to \ell^+ \, \ell^- + X$ at the center-of-mass energy $\sqrt{s} = 13$ TeV in terms of the $Z_R$ mass on the left panel, and coupling $g_R$ on the right panel. The plot on the left panel shows the limits obtained from ATLAS data analysis on dileptons extracted from Figure 3 in reference \cite{ATLAS:2019erb} as the full dark blue line, in comparison with the calculated cross section with $g_R = 0.55$, $g_{R}=0.65$ and $g_{R}=1.0$ as the magenta, green and orange lines, respectively. The dotted dark blue line is the conservatively extrapolated limit obtained by keeping the ATLAS cross section constant. The plot on the right panel shows the cross section as a function of the coupling $g_R$, with the masses $M_{Z_R} = 4$, $5$, and $6$ TeV plotted as the magenta, green, and orange lines, respectively.
    }
    \label{fig:ATLAS}
\end{figure}


We derive limits on the masses of left-right gauge bosons from dilepton searches involving a resonant neutral boson at the LHC. The leading-order cross section for the process $p \text{ } p \to Z_R \to \ell^+ \, \ell^- + X$ was computed using MadGraph \cite{Alwall:2014hca}, with the model described in section \ref{sec:model} implemented in FeynRules \cite{Alloul:2013bka}. To this end, we developed a tree-level algorithm based on the FeynRules implementation presented in reference \cite{Kriewald:2024cgr}. We emphasize that all observables were calculated at leading order, under the assumption that $\epsilon = v/v_R$ is small.\footnote{At next-leading order, a k-factor of 1.4 is observed in the production cross-section \cite{Kriewald:2024cgr}. Hence, our results are conservative.}, and an estimated error of the order of $\epsilon^2 \sim (\frac{246 \text{ GeV}}{5-10 \text{ TeV}})^2 \sim 10^{-3}$.

The most recent dilepton searches at the LHC include the ATLAS \cite{ATLAS:2019erb} and CMS \cite{CMS:2021ctt} data from Run II at $\sqrt{s} = 13$ TeV. The figure \ref{fig:ATLAS} compares the minimal LRSM results with the cross sections $p\text{ } p \to Z_R \to  \ell^+ \, \ell^- + X$ from ATLAS extracted from Figure 3 in reference \cite{ATLAS:2019erb,hepdata.88425}. 
The figure \ref{fig:ATLAS} shows benchmark curves calculated using $g_R = 0.55$, $0.65$, and $1.0$, represented by the magenta, green, and orange lines, respectively. The full dark blue curve represents the 95$\%$ C.L. curve extracted from ATLAS data, considering $\Gamma/M_{Z_R} = 3\%$. The ATLAS data is at most $M_{Z_R} = 6$ TeV. We have conservatively extrapolated this limit by keeping the cross section constant at higher mediator masses, as shown by the dotted dark blue line. The cross-section calculation using Madgraph also considered cuts on pseudo-rapidity $|\eta| < 2.5$ and 
 $p_T > 30$ GeV, as described by the ATLAS Collaboration in the reference \cite{ATLAS:2019erb}, and the parton distribution function sets LHAPDF \cite{Buckley:2014ana} 6.5.4 and pdfsets ``NNPDF40 nlo as 0119'', with  lhaid chosen as 334100 \cite{NNPDF:2021njg}. 
The figure \ref{fig:ATLAS} also shows on the right panel the cross section in terms of the coupling $g_R$, with the $Z_{R}$ masses of $M_{Z_R} = 4$ TeV in magenta, $5$ TeV green, and $6$ TeV orange lines.
The curve behavior is explained by observing the $Z_R$ interaction with fermions. The gauge boson has a larger production at low values of the coupling: $g_R \gtrsim g_L \, \tan(\theta_W) = 0.35$, as shown in the equation \ref{eq:gzr}.  Furthermore, the model is inconsistent with $g_R < 0.35$ because $v_R$ becomes imaginary. 
\begin{eqnarray}
v_R   = \frac{M_{Z_R}}{  g_R} \, \sqrt{1-\frac{g_L^2}{g_R^2} \, \tan^2(\theta_W)} \; .
\end{eqnarray}
Considering the $g_{R}$-coupling in the range of $0.35 < g_R < 1$, the ATLAS bounds constrain the VEV right-scale to $v_R \gtrsim 5 - 10$ TeV. 

\begin{figure*}
\vspace*{.5cm}       
\centering \resizebox{0.7\textwidth}{!}{%
  \includegraphics{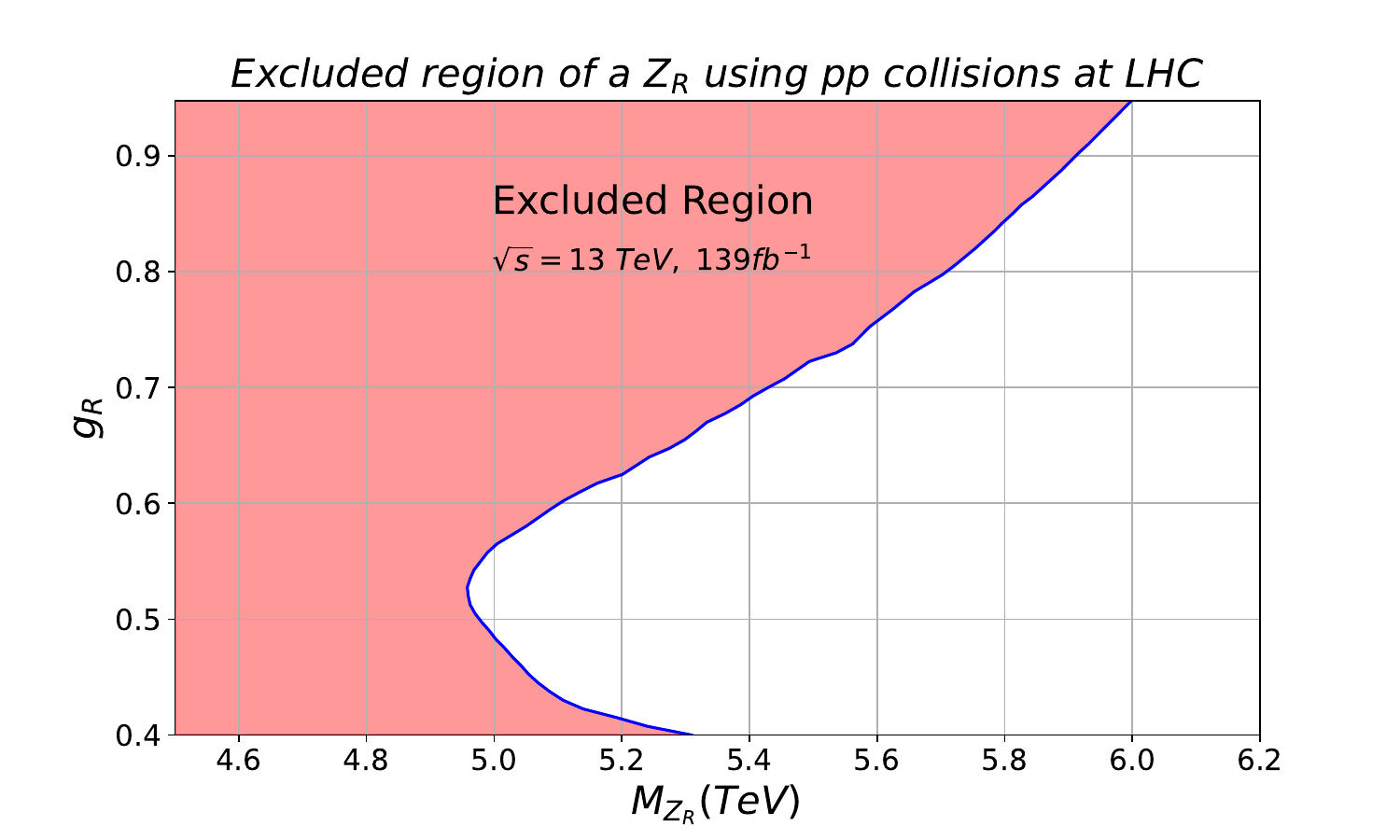}
}
    \caption{Exclusion region of the parameter space derived from $95\%$ C.L. dilepton cross section from ATLAS at $\sqrt{s}= 13$ TeV \cite{ATLAS:2019erb}. The collaboration data range from $M_{Z_R} = 1$ to 6 TeV; we have extrapolated the cross section  to derive the bounds at $M_{Z_R}> 6$ TeV.}
    \label{fig:95CL}
\end{figure*}
The figure \ref{fig:95CL} shows limits on the parameter space of the coupling $g_R$ and mass $M_{Z_R}$, constrained from $95\%$ C.L. from ATLAS data. For instance, $M_{Z_R} > 5.4 $ TeV at the benchmark point $g_R = g_L = 0.65$.  However, we obtained the excluded region with limits that vary from $M_{Z_R} > 4.9$ TeV up to $M_{Z_R} > 6.1$ TeV, depending on the coupling value $g_R$. The mass limit dependence on the coupling has a minimum value at $g_R$ = 0.55, which is also observed at the production cross section
 shown in Figure \ref{fig:ATLAS} on the right side. Notice that this is direct consequence of the denominator in equation \ref{eqzprimeint}.  A summary is included in  Table~\ref{tab:bounds}, where we compare the updated limits at $139$ fb$^{-1}$ with the previous results derived from LHC data at $3.2$ fb$^{-1}$ \cite{Lindner:2016lpp}.
\begin{table}[h!]
    \caption{$M_{Z_R}$ lower bounds from Run II at 13 TeV, with 3.2 fb$^{-1}$ \cite{Lindner:2016lpp} and 139 fb$^{-1}$ integrated luminosity calculated at the benchmark points $g_R = g_L$, and $g_R =1$. }
    \label{tab:bounds}   
    \centering
    \begin{tabular}{lll}
        \hline\noalign{\smallskip}
          & 3.2 fb$^{-1}$ & 139 fb$^{-1}$ \\
        \noalign{\smallskip}\hline\noalign{\smallskip}
               $g_R = g_L$  & $ 3$ TeV & $ 5.4$ TeV \\
               $g_R = 1$  & $ 4$ TeV & $ 6.1$ TeV \\
        \noalign{\smallskip}\hline
    \end{tabular}
\end{table}
%

%
\begin{figure}
\resizebox{0.5\textwidth}{!}{%
  \includegraphics{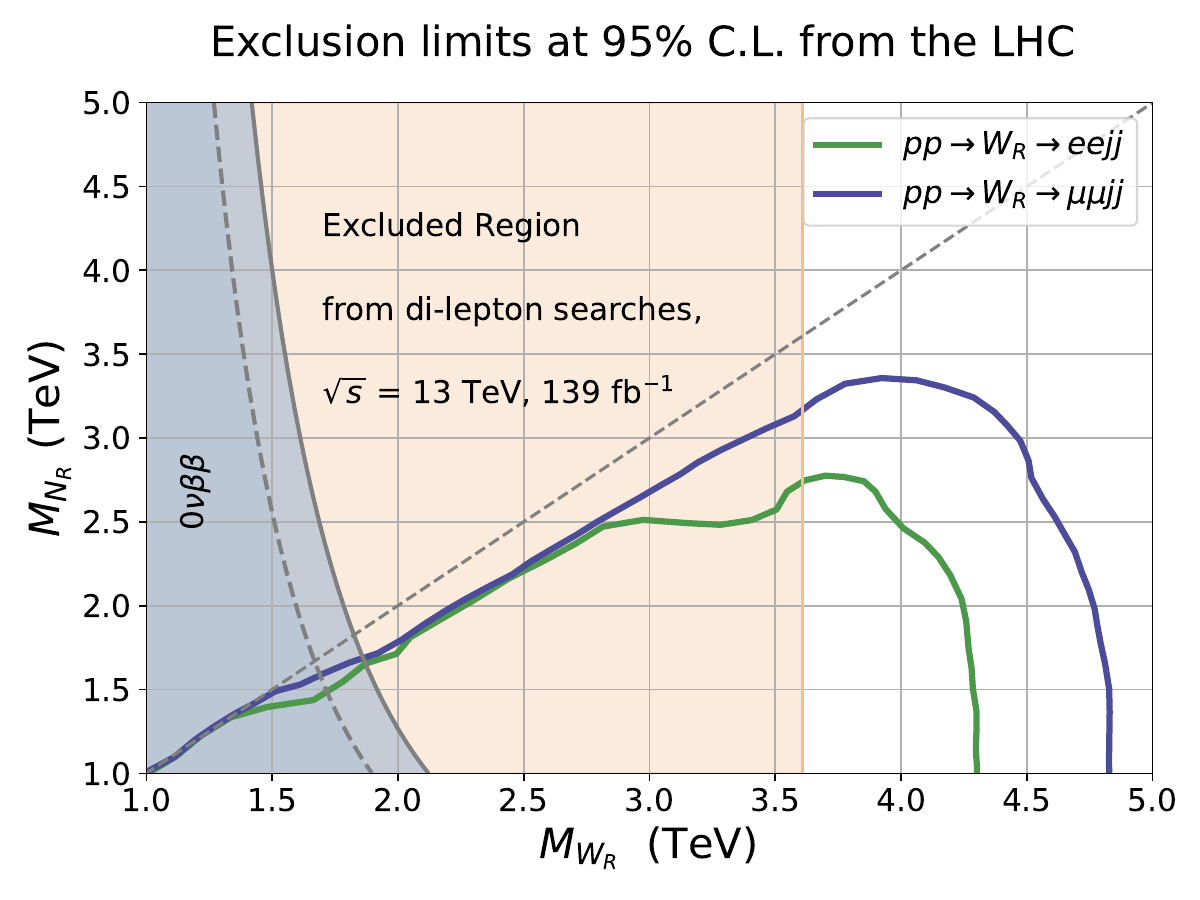}
}

    \caption{Dilepton bounds on the $M_{W_R} \times M_{N_R}$ mass plane. As the gauge boson masses are related (equation \eqref{eq:mwrmzr}) the dilepton bound (orange) represents an orthogonal and independent limit on the $W_R$ mass, namely $M_{W_R}> 3.18$ TeV for $g_R = g_L = 0.65$.The excluded region in orange is compared with the $95\%$ C.L. limits from the $p p \to W_R \to \ell\ell jj$ search channel, represented by the green ($eejj$) and blue ($\mu\mu jj$) lines extracted from \cite{Frank:2023epx}.  The gray regions indicate constraints from neutrinoless double beta decay; solid and dashed lines represent the KamLAND-Zen and GERDA limits, respectively}.
    \label{fig:MWRMN}
\end{figure}
These limits obtained in Figure \ref{fig:95CL} also influence direct searches for the boson $W_R$, since the gauge boson masses described in the equation \ref{eq:BosonsMasses} are related as: 
\begin{eqnarray}
\frac{ M_{W_R}}{M_{Z_R}} \simeq \sqrt{ \, \frac{1}{2}-\frac{g_L^2}{2g_R^2} \, \tan^2(\theta_W) } \; ,
\label{eq:mwrmzr}
\end{eqnarray}
which reads $M_{W_R} \approx 0.58 M_{Z_R}$.
Therefore, the limit on $g_R = g_L$ is $M_{Z_R} > 5.4$ TeV, implies the exclusion limit $M_{W_R} > 3.2$ TeV. Therefore, the dilepton searches offer a complementarity bound on the LRSM. For concreteness, we overlay our bounds on the $M_{W_R} \times M_{N_R}$ parameter space in Figure \ref{fig:MWRMN}.

In the Figure \ref{fig:MWRMN},  we exhibit the limits stemming from $W_R$ searches \cite{Frank:2023epx} and ours, for $g_R = g_L$. The orange region shows the dilepton limit $M_{W_R} > 3.2$ TeV, which covers the unexplored region of parameter space, where the right-handed neutrino is heavier than the gauge boson.
 Figure \ref{fig:MWRMN} also shows constraints from neutrinoless double beta decay ($0\nu\beta\beta$) . The contribution of the LRSM in this searches are discussed in detail in reference \cite{Patra:2023ltl}. The $0\nu\beta\beta$ half-life is calculated as
\begin{equation}
 \left[T_{1/2}^{0\nu}\right]^{-1} = (g_A^f)^4 \,G^{0\nu}_{01}~|{\cal M}^{0\nu}_\nu|^2 ~ \left| m_p \left(\frac{M_{W_L}}{M_{W_R}}\right )^4 \sum_{j=1,2,3} \frac{{\text{V}_{ej}}^2}{m_{N_{Rj}}}\right|^2,
\label{std_halflife}
\end{equation}
where $g_A^f =2\,( \, g_{L}^{\, f} - g_{R}^{\, f} \,)$ is the axial coupling and $g_{L(R)}^f$ defined in equation (\ref{eq:gf}), $G^{0\nu}_{01}$ is the phase space factor, ${\cal M}^{0\nu}_N$ represents the Nuclear Matrix Elements (NME), $m_p = 938.27$ MeV is the proton mass, and $M_{W_{L,R}}$ denote the $W$ boson masses. In this framework, the mixing $V_{ej}$ is approximately taken as the $U_{PMNS}$ matrix, and $m_{N_{Rj}}$ represents the right-handed neutrino mass. 

The values of the phase space factor and the NME are experiment-dependent; specifically, $G^{0\nu}_{01}$ is $0.22 \times 10^{-14} \text{ yr}^{-1}$ for Ge and $1.5 \times 10^{-14} \text{ yr}^{-1}$ for Xe. According to Table 3 of reference \cite{Patra:2023ltl}, the Nuclear Matrix Elements are ${\cal M}^{0\nu}_N = 104 - 401.3$ for Ge and ${\cal M}^{0\nu}_N = 66.9 - 186.3$ for Xe. Currently, the most stringent constraints on the half-life come from KamLAND-Zen ($^{136}$Xe) with $T_{1/2}^{0\nu} > 2.0 \times 10^{26}$ yr \cite{KamLAND-Zen:2024eml,KamLAND-Zen:2022tow,KamLAND-Zen:2016pfg} and GERDA ($^{76}$Ge) with $T_{1/2}^{0\nu} > 1.2 \times 10^{26}$ yr \cite{Agostini:2017iyd}.

Obviously, the precise relation between the masses of the $Z_R$ and $W_R$ bosons depends on the scalar sector. However, our qualitative conclusion holds. Searches for the $Z_R$ boson are worthwhile because they offer an orthogonal and independent view on the realization of the left-right symmetry in nature.


\section{Conclusions}
\label{conc}
%

The observation of charged right-handed current at the LHC represents a distinct signature of the left-right symmetry. Once a right-handed current is observed, the next step is to search for a dilepton resonance, as investigated in this work. For $g_R=g_L$, $M_{W_R}=0.58 M_{Z_{R}}$, thus the $W_R$ boson is lighter than the $Z_R$ boson. Thus, we naively expect to see first a signal of a $W_R$ boson. However, when the right-handed neutrino mass nears the $W_R$ boson, the LHC sensitivity to a $W_R$ signal weakens because the final leptons have a soft transverse momentum. For this reason, we believe it is worthwhile to conduct an independent probe for the $Z_R$ boson.

That said, we obtain updated constraints in the left-right symmetric model (LRSM) based on the resonant process $p\,p \rightarrow Z_{R} \rightarrow \ell^{+} \, \ell^{-}+X$ in the Run II at LHC, using dilepton data from the ATLAS collaboration at $\sqrt{s} = 13\ \text{TeV}$ of center-of-mass energy and $139\ \text{fb}^{-1}$ of integrated luminosity. Instead of focusing on a specific benchmark point  $g_R = g_L $, we varied $g_R$ and repeated this exercise to derive lower mass limits for different choices of $g_R$ (see figure \ref{fig:95CL}).  We found that searches for a $Z_R$ boson are complementary and cover an unexplored region of parameter space in the $M_{N_R} \geq M_{W_R}$.  

Therefore, the exciting hunt for a collider signature of the Left-Right symmetry should also rely on the search for a $Z_R$ boson.

\section{Acknowledgements}
M. J. Neves thanks the International Institute of Physics at the Federal University of Rio Grande do Norte (UFRN) for the kind and warm hospitality during his visit. We also thank Yohan Mauricio Oviedo-Torres for the discussions and recommendations. This work was funded by a CAPES Grant (88887.966347/2024-00).
FSQ is a Simons Foundation grantee (Award Number:1023171-RC). FSQ acknowledges support from CNPQ Grants 403521/2024-6, 408295/2021-0, 403521/2024-6, 406919/2025-9, 351851/ 2025-9, the FAPESP Grants 2021/01089-1, 2023/01197-4, ICTP-SAIFR 2021/14335-0, and the ANID-Millennium Science Initiative Program ICN2019\_044. This work is partially funded by FINEP under project 213/2024 and was carried out through the IIP cluster {\it bulletcluster}.

\pagebreak

\appendix
\section{The diagonalization of the gauge bosons sector}
\label{secAppendix}

In this appendix, we present some explicit results of the LRSM.
After the SSB mechanism, the charged gauge boson sector has a massive Lagrangian:
\begin{align}\label{LmWs}
{\cal L}_{mass}^{W}&=\frac{g_L^2 \, v_R^2}{4}
    \begin{pmatrix}
A_L^- & A_R^-
\end{pmatrix} \nonumber \\
&
\begin{pmatrix}
 \epsilon^2  & - \, e^{-i \alpha} \,  z_{LR} \, \epsilon^2 \, \sin{2 \beta} \\
- \, e^{i \alpha} \, z_{LR} \, \epsilon^2 \, \sin{2 \beta}  & \epsilon^2 + 2 \, z_{LR}^{2}
\end{pmatrix}
\begin{pmatrix}
A_L^+ \\ A_R^+
\end{pmatrix} \; ,
\end{align}
where $A_{L,R}^{\pm}=(A_{L,R}^{1}\mp i \, A_{L,R}^{2})/\sqrt{2}$, we have used the parameterization $\kappa_1=v\,\sin\beta$ and $\kappa_2=v\,\cos\beta$, with $\epsilon \equiv v/v_R$ and $z_{LR}\equiv g_R/g_L$. The mass matrix of (\ref{LmWs}) can be diagonalized by the transformation
\begin{align}\label{transfW}
\begin{pmatrix}
A_L^+ \\ A_R^+
\end{pmatrix} &=
\frac{1}{2 z_{LR}} \times  \nonumber \\ &
\begin{pmatrix}
\sqrt{4 z_{LR}^2 - \epsilon^2 \, \sin{2 \beta}} & \epsilon^2 \, \sin{2 \beta} \, e^{i \alpha} \\
- \, \epsilon^2 \, \sin{2 \beta} \, e^{-i\alpha}& \sqrt{4 z_{LR}^2 - \epsilon^2 \, \sin{2 \beta}}  
\end{pmatrix}
\begin{pmatrix}
W_L^+ \\ W_R^+
\end{pmatrix} \; ,
\end{align}
in which $W_{L,R}^{\pm}$ are the physical charged gauge bosons. The eigenvalues of the diagonalized matrix in (\ref{LmWs}) are given by  
\begin{eqnarray}
M_{W_L} \simeq \frac{g_L \, v}{2}
\hspace{0.5cm} \mbox{and} \hspace{0.5cm}
M_{W_R} \simeq \frac{g_R \, v_R}{\sqrt{2}} \; ,
\end{eqnarray}
where we neglected terms of order $\epsilon^2$, the mass $M_{W_L}$ is identified as the usual $W^{\pm}$ from the SM, and $M_{W_R}$ is the heavy charged gauge boson from LRSM.  
The neutral gauge bosons (\ref{LmassZ}) contains the mass matrix
\begin{equation}\label{MatrixMassa}
M^{2}=
\left(
\begin{array}{ccc}
v_{L}^{2} \, g^{2} + \frac{v^2}{4} \, g^2 & -\frac{v^2}{4} \, g^{2} & -g \, g_{BL} \, v_{L}^2
\\
\\
-\frac{v^2}{4} \, g^{2} & v_{R}^{2} \, g^{2}+\frac{v^2\,g^{2}}{4} & -g\,g_{BL}\,v_{R}^2
\\
\\
-g\,g_{BL}\,v_{L}^2 & -g\,g_{BL}\,v_{R}^2 & (v_L^2  + v_{R}^2) \, g_{BL}^{2}
\end{array}
\right) \; .
\end{equation}
It can be diagonalized by the $SO(3)$ transformations
\begin{subequations}\label{transfZ}
\begin{align}
A_{L\mu}^{\,\,3} &=\cos\theta_{W}\,\cos\theta_{3}\,Z_{\mu}+\cos\theta_{W}\,\sin\theta_{3}\,Z_{R\mu } + \nonumber \\ &
+\sin\theta_{W}\,A_{\mu} \; ,
\\
A_{R\mu}^{\,\,3} &= -\left(\sin\theta_{1}\,\sin\theta_{W}\,\cos\theta_{3}+\cos\theta_{1}\,\sin\theta_3\right)Z_{\mu}
\nonumber \\
&
\hspace{-0.6cm}
+\left(\sin\theta_{1}\,\sin\theta_{W}\,\cos\theta_{3}+\cos\theta_{1}\,\sin\theta_3\right)Z_{R\mu} + \nonumber \\ &+\sin\theta_{1}\,\cos\theta_{W}\,A_{\mu} \; ,
\\
B_{\mu} &= \left(-\cos\theta_{1}\,\sin\theta_{W}\,\cos\theta_3+\sin\theta_1\,\sin\theta_{3} \right) Z_{\mu}
\nonumber \\ 
&
\hspace{-0.6cm}
+\left(-\sin\theta_{1}\,\cos\theta_{3}-\cos\theta_{1}\,\sin\theta_{W}\,\sin\theta_{3}\right) Z_{R\mu}
+ \nonumber \\ &
+\cos\theta_{1}\,\cos\theta_{W}\,A_{\mu} \; ,
\end{align}
\end{subequations}
where the $\theta_{1}$-mixing angle is $\sin\theta_1=(g_L/g_{R})\tan\theta_{W}$, for $g_L \neq g_R$, and the $\theta_{3}$ is 
\begin{align}
\tan(2\theta_{3})&\simeq -\frac{g_R^2\,v^2}{2(g_{R}^2+g_{BL}^2)^2 \, v_{R}^2} 
\nonumber \\ 
&
\times \sqrt{ g_L^2\,g_{R}^2+g_{L}^2\,g_{BL}^2+g_{R}^2\,g_{BL}^2  } \; ,    
\end{align}
that lead the relations (\ref{theta1theta3}). The new basis $(Z_{\mu},Z_{R\mu},A_{\mu})$ are the physical 
neutral gauge bosons whose mass eigenstates 
are the eigenvalues of the mass matrix (\ref{MatrixMassa}) :  
\begin{subequations}
\begin{eqnarray}
M_{Z}^2 &\simeq& \frac{g_{L}^2\,g_{R}^2+g_{L}^2\,g_{BL}^2+g_{R}^2\,g_{BL}^2}{4\,(g_{R}^2+g_{BL}^2)} \, v^2 =\frac{g_{L}^2 \, v^2}{4\,\cos^2\theta_{W}}  \; , \; \; \; \; \; \;
\\
M_{Z_{R}}^2 &\simeq& (g_{R}^2+g_{BL}^2) \, v_{R}^2=\frac{g_{R}^2 \, v_{R}^2 }{1-(g_{L}^2/g_{R}^2)\tan^2\theta_{W}} \; , \; \;
\end{eqnarray}
\end{subequations}
and $M_{A}=0$ for the photon mass. Using the transformations (\ref{transfW}) and (\ref{transfZ}) in the covariant derivative operator (\ref{Dmu}) acting on the fermions in (\ref{Lfermions}), the new couplings of $W_{R}$ and $Z_{R}$ with fermions are   
\begin{eqnarray}
{\cal L}^{int}_{W_{R}-f}=\frac{g_R}{\sqrt2} \, W_{R\mu}^{+} \left[ \, \overline{u}_R \, \gamma^\mu \, d_R + \overline{N}_R \, \gamma^\mu \, \ell_R \, \right] + {\rm h.c.} \; , \;\;
\end{eqnarray}
and 
\begin{align}
{\cal L}^{int}_{Z_{R}-f} &= \frac{g_R}{\sqrt{1 - (g_L^2 / g_R^2) \, \tan^2 \theta_W }} \, Z_{R\mu} \, 
\overline{f} \, \gamma^\mu \nonumber \\ &
\times \, \left[ \, I_{3R} + \frac{g_L^2}{g_R^2} \, \tan^2 \theta_W \,
(I_{3L} - Q_{em}) \, \right] f \; .
\end{align} 

We do not specify the scalar sector of this model; for a more comprehensive study, one may refer to references \cite{Solera:2023kwt,Kriewald:2024cgr,Dev:2016dja,Maiezza:2016ybz}.


\begin{thebibliography}{}
%
%
\bibitem{PatiPRD1974} J.C. Pati and A. Salam, {\it Lepton number as the fourth color}, Phys. Rev. D, {\bf 10} 275 (1974).


\bibitem{MohapatraPRD1975} R. N. Mohapatra and J. C. Pati, {\it Natural left-right symmetry}, Phys. Rev. D, {\bf 11} 2558 (1975).

\bibitem{PhysRevD.11.566},
Mohapatra, Rabindra N. and Pati, Jogesh C.,
{\it Left-right gauge symmetry and an "isoconjugate" model of $\mathrm{CP}$ violation},
Phys. Rev. D, {\bf 11}, 3 (1975), 10.1103/PhysRevD.11.566,
https://link.aps.org/doi/10.1103/PhysRevD.11.566



\bibitem{SenjanoviPRD1975} G. Senjanovic and R. N. Mohapatra, {\it Exact left-right symmetry and spontaneous violation of parity}, Phys. Rev. D, {\bf 12} 1502 (1975).


\bibitem{seesaw1} P. Minkowski, {\it $\mu \rightarrow e\, \gamma$ at a rate of one out of $10^{9}$ muon decays ?}, Phys. Lett. B, {\bf 67} 421 (1977).


\bibitem{seesaw2} R. N. Mohapatra and G. Senjanovi\'{c}, {\it Neutrino Mass and Spontaneous Parity Nonconservation}, Phys. Rev. Lett. {\bf 44}, 912 (1980).


\bibitem{seesaw3} T. Yanagida, {\it Horizontal gauge symmetry and masses of neutrinos}, Conf.  Proc.  C {\bf 7902131},  95  (1979).


\bibitem{seesaw4} M. Gell-Mann, P. Ramond and R. Slansky, {\it Complex Spinors and Unified Theories}, 
Conf. Proc. C {\bf 790927}, 315 (1979) [arXiv:1306.4669 [hep-th]].


\bibitem{seesaw5} S.~L.~Glashow, {\it The Future of Elementary Particle Physics}, NATO Sci. Ser. B {\bf 61}, 687 (1980).


\bibitem{MJNevesHelayelMohapatraOkada2018} M. J. Neves, J. A. Hela\"yel-Neto, R. N. Mohapatra and N. Okada,
{\it Minimally Extended Left-Right Symmetric Model for Dark Matter with $U(1)$ Portal}, JHEP \textbf{12}, 009 (2018)
[arXiv:1808.00484 [hep-ph]].





\bibitem{MJNevesOkada2021} M. J. Neves, S. Okada and N. Okada,
{\it Majorana fermion dark matter in minimally extended left-right symmetric model}, JHEP \textbf{09}, 038 (2021)
[arXiv:2103.08873 [hep-ph]].




\bibitem{Fuks:2025jrn}
B.~Fuks, J.~Kriewald, M.~Nemev{\v{s}}ek and F.~Nesti,
{\it ``Beautiful Majorana Higgses at colliders,''}
JHEP \textbf{06} (2025), 254
doi:10.1007/JHEP06(2025)254
[arXiv:2503.21354 [hep-ph]].

\bibitem{ThomasArun:2021rwf}
M.~Thomas Arun, T.~Mandal, S.~Mitra, A.~Mukherjee, L.~Priya and A.~Sampath,
Phys. Rev. D \textbf{105} (2022) no.11, 115007
doi:10.1103/PhysRevD.105.115007
[arXiv:2109.09585 [hep-ph]].

\bibitem{Hong:2023mwr}
T.~T.~Hong, V.~K.~Le, L.~T.~T.~Phuong, N.~C.~Hoi, N.~T.~K.~Ngan and N.~H.~T.~Nha,
{\it ``Decays of Standard Model{\textendash}Like Higgs Boson $h \rightarrow\gamma\gamma, Z \gamma$in a Minimal Left-Right Symmetric Model,''}
PTEP \textbf{2024} (2024) no.3, 033B04
[erratum: PTEP \textbf{2024} (2024) no.5, 059201]
doi:10.1093/ptep/ptae029
[arXiv:2312.11045 [hep-ph]].



\bibitem{Liu:2025ldf}
W.~Liu and Z.~Chen,
{\it ``Probing the Left-Right Symmetric Model from Displaced Shower at CMS Muon System,''}
[arXiv:2503.22112 [hep-ph]].

\bibitem{Lindner:2016lpp}
M.~Lindner, F.~S.~Queiroz and W.~Rodejohann,
{\it ``Dilepton bounds on left\textendash{}right symmetry at the LHC run II and neutrinoless double beta decay,''}
Phys. Lett. B \textbf{762} (2016), 190-195
doi:10.1016/j.physletb.2016.08.068
[arXiv:1604.07419 [hep-ph]].

\bibitem{De:2024puh}
S.~De, A.~Dey and T.~Samui,
{\it ``Jet substructure analysis for distinguishing left- and right-handed couplings of heavy neutrino in W' decay at the HL-LHC,''}
Phys. Rev. D \textbf{111} (2025) no.7, 075017
doi:10.1103/PhysRevD.111.075017
[arXiv:2411.14910 [hep-ph]].

%
\bibitem{Nemevsek:2011hz} Miha Nemevsek, Fabrizio Nesti, Goran Senjanovic and Yue Zhang, 
{\it First Limits on Left-Right Symmetry Scale from LHC Data}, Phys. Rev. D, {\bf 83} 115014 (2011).




\bibitem{Nemevsek:2023hwx} Miha Nemev\v{s}ek and Fabrizio Nesti, {\it Left-right symmetry at an FCC-hh}, Phys. Rev. D, {\bf 108} 015030 (2023).


\bibitem{Zhang:2007da}
Y.~Zhang, H.~An, X.~Ji and R.~N.~Mohapatra,
{\it ``General CP Violation in Minimal Left-Right Symmetric Model and Constraints on the Right-Handed Scale''}, 
Nucl. Phys. B \textbf{802} (2008), 247-279
doi:10.1016/j.nuclphysb.2008.05.019
[arXiv:0712.4218 [hep-ph]].

\bibitem{Bajc:2009ft}
B.~Bajc, M.~Nemevsek and G.~Senjanovic,
Phys. Lett. B \textbf{684} (2010), 231-235
doi:10.1016/j.physletb.2010.01.025
[arXiv:0911.1323 [hep-ph]].

\bibitem{Alves:2022yav}
G.~F.~S.~Alves, C.~S.~Fong, L.~P.~S.~Leal and R.~Z.~Funchal,
{\it ``Exploring the Neutrino Sector of the Minimal Left-Right Symmetric Model,''}
[arXiv:2208.07378 [hep-ph]].



\bibitem{Maiezza:2010ic}
A.~Maiezza, M.~Nemevsek, F.~Nesti and G.~Senjanovic,
{\it ``Left-Right Symmetry at LHC''}, 
Phys. Rev. D \textbf{82} (2010), 055022
doi:10.1103/PhysRevD.82.055022
[arXiv:1005.5160 [hep-ph]].


\bibitem{Frank:2023epx}
M.~Frank, B.~Fuks, A.~Jueid, S.~Moretti and O.~Ozdal,
{\it ``A novel search strategy for right-handed charged gauge bosons at the Large Hadron Collider''}, 
JHEP \textbf{02} (2024), 150
doi:10.1007/JHEP02(2024)150
[arXiv:2312.08521 [hep-ph]].

\bibitem{Patra:2015bga}
S.~Patra, F.~S.~Queiroz and W.~Rodejohann,
Phys. Lett. B \textbf{752} (2016), 186-190
doi:10.1016/j.physletb.2015.11.009
[arXiv:1506.03456 [hep-ph]].





\bibitem{Osland:2020onj}
P.~Osland, A.~A.~Pankov and I.~A.~Serenkova,
{\it ``Updated constraints on $Z'$ and $W'$ bosons decaying into bosonic and leptonic final states using the run 2 ATLAS data''}, 
Phys. Rev. D \textbf{103} (2021) no.5, 053009
doi:10.1103/PhysRevD.103.053009
[arXiv:2012.13930 [hep-ph]].

\bibitem{Solera:2023kwt}
S.~F.~Solera, A.~Pich and L.~Vale Silva,
{\it ``Direct bounds on Left-Right gauge boson masses at LHC Run 2 ''}, 
JHEP \textbf{02} (2024), 027
doi:10.1007/JHEP02(2024)027
[arXiv:2309.06094 [hep-ph]].
%
\bibitem{Kriewald:2024cgr} Jonathan Kriewald, Miha Nemev\v{s}ek, and Fabrizio Nesti, {\it Enabling precise predictions for left-right symmetry at colliders},
Eur. Phys. J. C, {\bf 84}, 1306 (2024).
%

\bibitem{Chakrabortty:2016wkl}
J.~Chakrabortty, J.~Gluza, T.~Jelinski and T.~Srivastava,
{\it ``Theoretical constraints on masses of heavy particles in Left-Right Symmetric Models''}, 
Phys. Lett. B \textbf{759} (2016), 361-368
doi:10.1016/j.physletb.2016.05.092
[arXiv:1604.06987 [hep-ph]].

\bibitem{Deppisch:2015cua}
F.~F.~Deppisch, L.~Graf, S.~Kulkarni, S.~Patra, W.~Rodejohann, N.~Sahu and U.~Sarkar,
{\it ``Reconciling the 2 TeV excesses at the LHC in a linear seesaw left-right model''}, 
Phys. Rev. D \textbf{93} (2016) no.1, 013011
doi:10.1103/PhysRevD.93.013011
[arXiv:1508.05940 [hep-ph]].


%
\bibitem{Lee:2017mfg} Chang Hun Lee, {\it Left-right symmetric model and its TeV-scale phenomenology}, http://drum.lib.umd.edu/handle/1903/20007 (2017).

\bibitem{ATLAS:2019erb}
G.~Aad \textit{et al.} [ATLAS],
{\it ``Search for high-mass dilepton resonances using 139 fb$^{-1}$ of $pp$ collision data collected at $\sqrt{s}=$13 TeV with the ATLAS detector,''}
Phys. Lett. B \textbf{796}, 68-87 (2019)
doi:10.1016/j.physletb.2019.07.016
[arXiv:1903.06248 [hep-ex]].

\bibitem{Alwall:2014hca}
J.~Alwall, R.~Frederix, S.~Frixione, V.~Hirschi, F.~Maltoni, O.~Mattelaer, H.~S.~Shao, T.~Stelzer, P.~Torrielli and M.~Zaro,
{\it ``The automated computation of tree-level and next-to-leading order differential cross sections, and their matching to parton shower simulations''}, 
JHEP \textbf{07} (2014), 079
doi:10.1007/JHEP07(2014)079
[arXiv:1405.0301 [hep-ph]].


\bibitem{Alloul:2013bka}
A.~Alloul, N.~D.~Christensen, C.~Degrande, C.~Duhr and B.~Fuks,
{\it ``FeynRules  2.0 - A complete toolbox for tree-level phenomenology,''}
Comput. Phys. Commun. \textbf{185} (2014), 2250-2300
doi:10.1016/j.cpc.2014.04.012
[arXiv:1310.1921 [hep-ph]].



\bibitem{CMS:2021ctt}
A.~M.~Sirunyan \textit{et al.} [CMS],
{\it ``Search for resonant and nonresonant new phenomena in high-mass dilepton final states at $ \sqrt{s} $ = 13 TeV,''}
JHEP \textbf{07} (2021), 208
doi:10.1007/JHEP07(2021)208
[arXiv:2103.02708 [hep-ex]].


\bibitem{hepdata.88425}
G.~Aad \textit{et al.} [ATLAS],
{\it ``Search for high-mass dilepton resonances using 139 fb$^{-1}$ of $pp$ collision data collected at $\sqrt{s}=$13 TeV with the ATLAS detector,''}
[ HEPData (collection), 2019]
{\it https://doi.org/10.17182/hepdata.88425} .


\bibitem{Buckley:2014ana}
A.~Buckley, J.~Ferrando, S.~Lloyd, K.~Nordstr\"om, B.~Page, M.~R\"ufenacht, M.~Sch\"onherr and G.~Watt,
{\it ``LHAPDF6: parton density access in the LHC precision era''}, 
Eur. Phys. J. C \textbf{75} (2015), 132
doi:10.1140/epjc/s10052-015-3318-8
[arXiv:1412.7420 [hep-ph]].

\bibitem{NNPDF:2021njg}
R.~D.~Ball \textit{et al.} [NNPDF],
{\it ``The path to proton structure at 1\% accuracy,''}
Eur. Phys. J. C \textbf{82} (2022) no.5, 428
doi:10.1140/epjc/s10052-022-10328-7
[arXiv:2109.02653 [hep-ph]].


\bibitem{ATLAS:2020zzb}
G.~Aad \textit{et al.} [ATLAS],
{\it``Search for dijet resonances in events with an isolated charged lepton using $\sqrt{s} = 13$ TeV proton-proton collision data collected by the ATLAS detector''},  
JHEP \textbf{06} (2020), 151
doi:10.1007/JHEP06(2020)151
[arXiv:2002.11325 [hep-ex]].

\bibitem{Patra:2023ltl}
S.~Patra, S.~T.~Petcov, P.~Pritimita and P.~Sahu,
{\it ``Neutrinoless double beta decay in a left-right symmetric model with a double seesaw mechanism,''}, 
Phys. Rev. D \textbf{107} (2023) no.7, 075037
doi:10.1103/PhysRevD.107.075037
[arXiv:2302.14538 [hep-ph]].


\bibitem{KamLAND-Zen:2024eml}
S.~Abe \textit{et al.} [KamLAND-Zen],
{\it ``Search for Majorana Neutrinos with the Complete KamLAND-Zen Dataset,''}, 
Phys. Rev. Lett. \textbf{135} (2025) no.26, 262501
doi:10.1103/jkf6-48j8
[arXiv:2406.11438 [hep-ex]].

\bibitem{KamLAND-Zen:2022tow}
S.~Abe \textit{et al.} [KamLAND-Zen],
{\it ``Search for the Majorana Nature of Neutrinos in the Inverted Mass Ordering Region with KamLAND-Zen,''
Phys. Rev. Lett. \textbf{130} (2023) no.5, 051801}, 
doi:10.1103/PhysRevLett.130.051801
[arXiv:2203.02139 [hep-ex]].

\bibitem{KamLAND-Zen:2016pfg}
A.~Gando \textit{et al.} [KamLAND-Zen],
{\it ``Search for Majorana Neutrinos near the Inverted Mass Hierarchy Region with KamLAND-Zen,''}, 
Phys. Rev. Lett. \textbf{117} (2016) no.8, 082503
doi:10.1103/PhysRevLett.117.082503
[arXiv:1605.02889 [hep-ex]].


\bibitem{Agostini:2017iyd}
M.~Agostini, M.~Allardt, A.~M.~Bakalyarov, M.~Balata, I.~Barabanov, L.~Baudis, C.~Bauer, E.~Bellotti, S.~Belogurov and S.~T.~Belyaev, \textit{et al.}
{\it ``Background-free search for neutrinoless double-$\beta$ decay of $^{76}$Ge with GERDA,''}, 
Nature \textbf{544} (2017), 47
doi:10.1038/nature21717
[arXiv:1703.00570 [nucl-ex]].

\bibitem{Dev:2016dja}
P.~S.~B.~Dev, R.~N.~Mohapatra and Y.~Zhang,
{\it ``Probing the Higgs Sector of the Minimal Left-Right Symmetric Model at Future Hadron Colliders,''}, 
JHEP \textbf{05} (2016), 174
doi:10.1007/JHEP05(2016)174
[arXiv:1602.05947 [hep-ph]].


\bibitem{Maiezza:2016ybz}
A.~Maiezza, G.~Senjanovi{\'c} and J.~C.~Vasquez,
{\it ``Higgs sector of the minimal left-right symmetric theory,''}, 
Phys. Rev. D \textbf{95} (2017) no.9, 095004
doi:10.1103/PhysRevD.95.095004
[arXiv:1612.09146 [hep-ph]].

\end{thebibliography}
%

\end{document}